# Research on the laser – brain tissue interaction by finite element analysis

Xianlin Song [a, #, *], Ao Teng [b, #], Jianshuang Wei [c, d, #], Lingfang Song [e]
[a] School of Information Engineering, Nanchang University, Nanchang 330031, China;
[b] Ji luan Academy, Nanchang University, Nanchang 330031, China;
[c] Britton Chance Center for Biomedical Photonics, Wuhan National Laboratory for Optoelectronics-Huazhong University of Science and Technology, Wuhan 430074, China;
[d] Moe Key Laboratory of Biomedical Photonics of Ministry of Education, Department of Biomedical Engineering, Huazhong University of Science and Technology, Wuhan 430074, China;
[e] Nanchang Normal University, Nanchang 330031, China;
# equally contributed to this work
* Corresponding author: songxianlin@ncu.edu.cn

## ABSTRACT

The study of the interaction between laser and brain tissue has important theoretical and practical significance for brain imaging. A two-dimensional simulation model that studies the propagation of light and heat transfer in brain tissue based on finite element has been developed by using the commercial finite element simulation software COMSOL Multiphysics. In this study, the model consists of three parts of 1) a layer of water on the surface of the brain, 2) brain tissue and 3) short pulsed laser source (the wavelength is 840nm). The laser point source is located in the middle of the layer of water above the brain tissue and irradiates the brain tissue. The propagation of light in brain tissue was simulated by solving the diffusion equation. And the temperature changes of gray matter and blood vessels were achieved by solving the biological heat transfer equation. The simulation results show that the light energy in the brain tissue decreases exponentially with the increase of penetration depth. Since the cerebral blood vessels have a stronger absorption on light compared with the surrounding tissues, the remaining light energy of the blood vessels in the cerebral cortex is ~ 85.8% of the remaining light energy in the surrounding gray matter. In the process of biological heat transfer, due to more light deposition in blood vessels, the temperature of blood vessels is 0.15 K higher than that of gray matter, and the temperature of gray matter hardly changes. This research is helpful to understand the propagation of light in the brain and the interaction between them, and has certain theoretical guiding for the optical imaging of the brain.

## 1. INTRODUCTION

The brain is the main part of the central nervous system including structures such as the cerebrum, cerebellum, and brainstem. The brain structure is intricate with hundreds of billions of neurons. There are many nerve centers composed of nerve cells and a large number of up and down nerve fiber bundles pass through and connect the cerebrum, cerebellum and spinal cord. Therefore, it is very important for brain imaging [1]. Laser has a wide range of applications in the brain imaging[2]. The theory of photon transmission based on light particles is the most widely used and most successful theory. The radiation transfer equation (RTE) accurately describes the propagation of light in the biological tissues, and the Monte Carlo (MC) simulation model provides an accurate solution for weak absorption and high scattering media [3]. Diffusion approximation (DA) is a typical method that uses RTE as a forward photon propagation model[4]. The method of diffusion approximation is widely used due to its fast calculation and high accuracy, and now it has become the basis of many basic applications based on the DA theory. A numerical simulation model composed of water, brain tissue and short-pulse laser source was developed by using the commercial finite element simulation software COMSOL Multiphysics. The model studied the propagation of laser and heat transfer in the brain tissue. The transfer of light energy from the water layer to the brain tissue is described by the diffusion equation or the Helmholtz equation. The temperature changes of gray matter and blood vessel are obtained by solving the biological heat transfer equation.

In this chapter, a two-dimensional rectangular model is used to simulate a partial cross-section of the brain. The diffusion equation or the Helmholtz equation describes the transmission of laser in the brain tissue. The biological heat transfer equation describes the temperature distribution of gray matter and blood vessel.

## 2. MODEL AND METHOD

### 2.1 Geometry

The brain is composed of the scalp, skull, subarachnoid space (filled with cerebrospinal fluid (CSF)), gray matter, white matter and many blood vessels. A two-dimensional simulation model based on finite element has been developed by using the commercial finite element simulation software COMSOL Multiphysics. For simplicity, the actual structure of the brain has been simplified as shown in Figure 1. The thickness of scalp, skull, cerebrospinal fluid and gray matter is 3 mm, 4 mm, 2 mm and 4 mm respectively. The diameter of blood vessel is 60 um and the length of it is 5 mm. There is a water layer on the scalp and its thickness is 1 mm. The pulsed laser point source is placed in the middle of the water layer to irradiate the brain tissue.

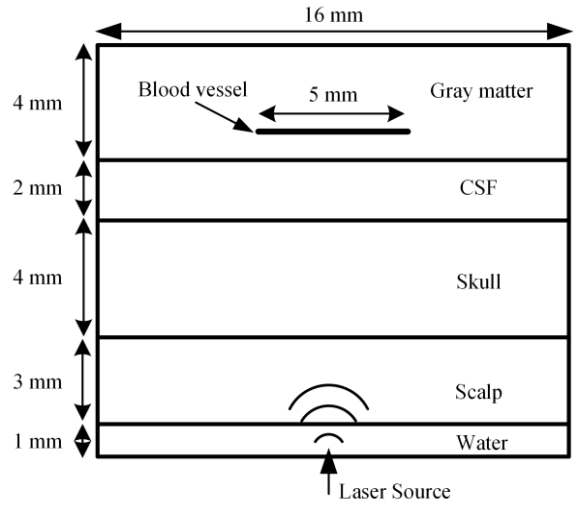

Figure 1. Model geometry

### 2.2 Light propagation

The propagation of laser in the brain tissue is simulated by using the partial differential module in the form of coefficient. In the process of simulating the transmission of laser through water and brain tissue to blood vessel, the laser source emits a Gaussian pulse with energy $W_p = 4\ mJ/cm^2$ and pulse width $\tau_p = 10\ ns$. This module uses the Helmholtz equation [5] given in Equation (1) to solve the fluence rate $\varphi$.

$$\frac{n}{c}\frac{\partial \varphi}{\partial t} + \nabla \cdot (-D\nabla \varphi) + u_a = f \quad (1)$$

$$f = \frac{1}{4\pi^{1.5}}\frac{W_p}{\tau_p}\exp(-\frac{4\cdot(t-\tau_c)^2}{\tau_p^2})\delta(r-r_0) \quad (2)$$

Where, $D = 1/(3(u_a + u_s'))$ is the diffusion coefficient, $u_s' = u_s(1-g)$ is the reduced scattering coefficient, $u_s$ is the scattering coefficient, $g$ is the anisotropic factor, $u_a$ is the absorption coefficient, $c$ is the speed of light in vacuum, $f$ represents the laser source at point $r_0$ with Gaussian distribution, which emits a laser pulse at $\tau_c = 30\ ns$. The optical parameters of brain tissue and water are listed in Table 1:

Table 1. Human adult brain model with optical properties.

|  | $u_a$ (mm$^{-1}$) | $u_s$ (mm$^{-1}$) |
| --- | --- | --- |
| Scalp | 0.021 | 1.81 |
| Skull | 0.019 | 1.52 |
| CSF | 0.005 | 0.23 |
| Gray matter | 0.042 | 2.09 |
| Blood vessel | 0.233 | 0.522 |
| Water | 0.006 | 0.001 |

## 2.3 Heat transfer

Under the irradiation of laser, there will be a certain regular light distribution in the brain tissue. Brain tissue absorbs laser energy and generates heat energy inside and on the surface of the brain tissue. Due to the absorption of light energy, the temperature in the brain tissue will change, which can be simulated by using the module of biological heat transfer in COMSOL Multiphysics to solve Equation (3):

$$\rho C \frac{\partial T}{\partial t} - \nabla \cdot (k \nabla T) = u_a \varphi \quad (3)$$

Where, $\rho$ is the density of the tissue, $C$ is the specific heat capacity, and $k$ is the thermal conductivity. Set the initial temperature of the tissue to 310.15 K. For simplicity, this model only simulates the temperature changes of gray matter and blood vessel. Since the moment of interaction of laser with brain tissue is short, the regulation mechanism of temperature is not considered. The thermal properties of gray matter and blood vessel are listed in Table 2:

Table 2. Thermal Properties.

|  | Density, $\rho$ (Kg·m$^3$) | Specific Heat Capacity, C (J/kg·K) | Thermal Conductivity, k (W/m·K) |
| --- | --- | --- | --- |
| Gray matter | 998 | 840 | 0.7 |
| Blood vessel | 1000 | 3639 | 1.1 |

## 3. RESULTS AND DISCUSSION

### 3.1 Light propagation, absorption and penetration depth

In the simulation, the emission time of laser pulse is 30 ns, and the width of pulse is 20 ns. The light energy distribution of the simulation model based on finite element analysis is shown in Figure 2. In order to better observe the propagation of light in the brain tissue, the water layer of 0 to 1 mm is hidden in the z-axis direction. Figures 2(a)-2(c) and Figures 2 (d) show the normalized light energy distribution in the brain tissue at four different moments. Figure 2. (a) shows that the laser source has not emitted laser pulse and the light energy in the brain tissue is zero at 20 ns. Figure 2(b) shows that the laser source has already emitted laser pulse at 26 ns, but the intensity of the laser pulse has not reached the peak. Figure 2(c) shows that the intensity of the laser pulse emitted by the laser source has reached the peak at 30 ns, and the light energy in the entire brain tissue has reached the maximum at this time. Figure 2(d) shows that the intensity of the laser pulse emitted by the laser source has been weakened at 34 ns, and the light energy in the brain tissue is decreasing.

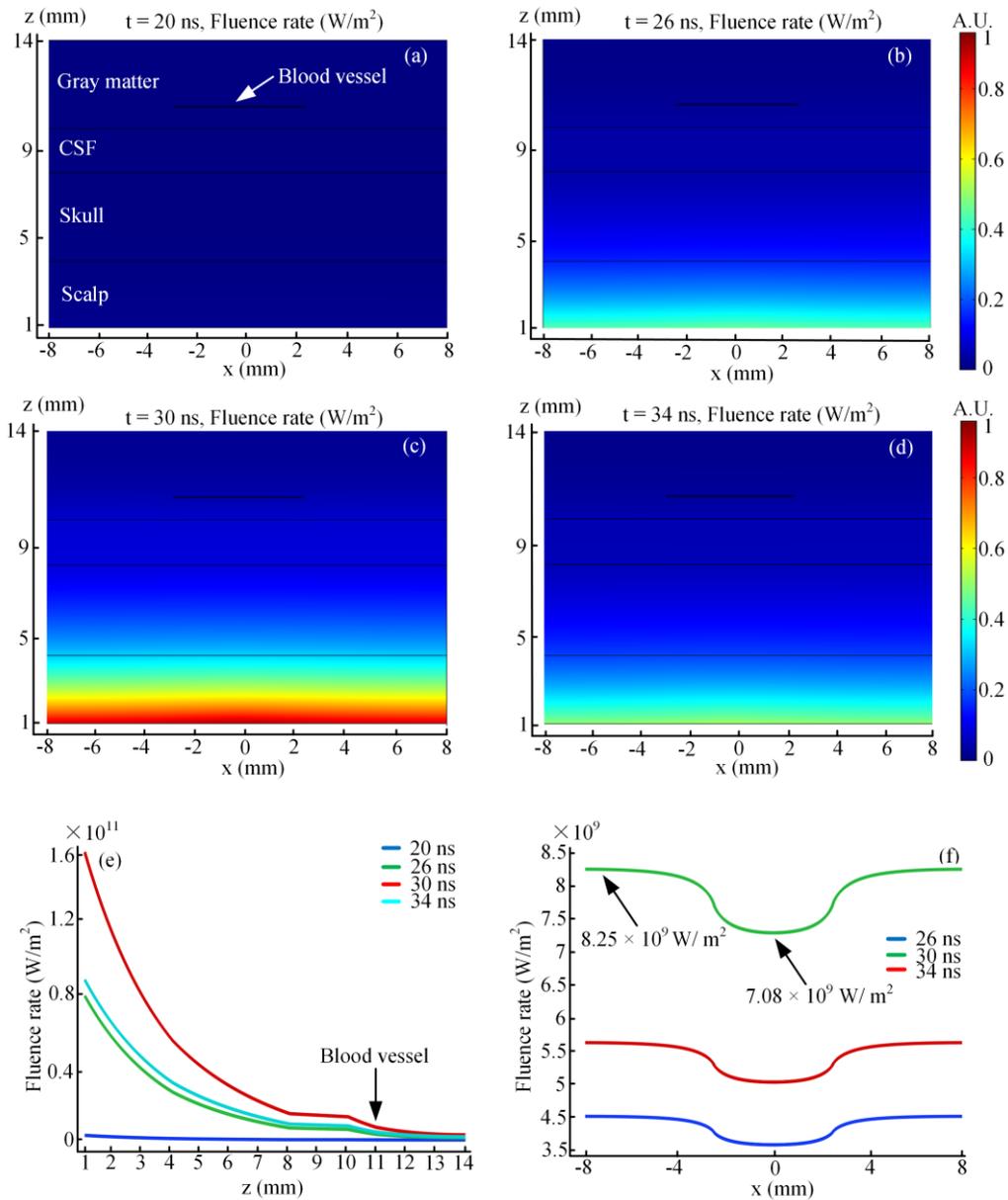

Figure 2. The normalized fluence rate of different moments at (a) 20 ns; (b) 26 ns; (c) 30 ns; (d) 34 ns. Figure 2. (e) shows the fluence rate of different moments at 20 ns, 26 ns, 30 ns, 34 ns of x = 0 mm. Figure 2. (f) shows the fluence rate of different moments at 26 ns, 30 ns, 34 ns of z = 11 mm. It also can be seen from Figure 2. (f) that at the moment of laser pulse emission ( t = 30 ns ), the remaining light energy inside the blood vessel is about $7.08 \times 10^9$ W/m², and the remaining light energy at the gray matter around the blood vessel is about $8.25 \times 10^9$ W/m².

It can also be seen that the light energy decreases as the penetration depth of laser increases from Figures 2(a)-2(c) and (d). In order to deeply understand the distribution of light energy, Figure 2(e) shows the light energy distribution of x = 0 mm at 20, 26, 30 and 34 ns. It can be seen that light energy decreases exponentially with the increase of the penetration depth of laser from Figure 2(e). It can also be seen that there are obvious mutations of light energy in cerebrospinal fluid and skull, cerebrospinal fluid and gray matter, because their optical parameters are different. In fact, there are mutations of the light energy in the skull and scalp, gray matter and blood vessel. The mutations are so unobvious that they are not shown intuitively in Figure 2(e). In order to intuitively understand the distribution of light energy in the gray matter and blood vessel, Figure 2(f) shows the light energy distribution of z = 11 mm through the blood vessel. It can be seen that at 26, 30 and 34 ns, the remaining light energy at the blood vessel is lower than that in the surrounding gray matter. This is because the absorption coefficient of the blood vessel is significantly higher than that of the gray matter, and so the blood vessel absorbs more photons resulting in the remaining light energy at the blood vessel less than the remaining light energy at the gray matter. At the moment of laser pulse emission (t = 30 ns), the remaining light energy inside the blood vessel is about $7.08 \times 10^9$ W/m$^2$, and the remaining light energy at the gray matter around the blood vessel is about $8.25 \times 10^9$ W/m$^2$, so the remaining light energy inside the blood vessel is about 85.8% of the remaining light energy of the surrounding gray matter. At the same time, the remaining light energy in the center of the blood vessel is slightly lower than that at both ends of the blood vessel.

### 3.2 Temperature distribution

It can be seen that the temperature of blood vessel is higher than that of the gray matter, and the temperature of gray matter is almost uniformly distributed from Figure 3(a). Figure 3(b) is an enlarged view of the rectangular dotted area in Figure 3(a), and it shows that the temperature of blood vessel is 0.15 K higher than that of the gray matter. In order to quantitatively understand the difference of temperature between blood vessel and gray matter, two temperature distribution diagrams of two lines were drawn. The temperature distribution of x = 0 mm through the blood vessel and gray matter is shown in Figure 3(c). Figure 3(d) shows the temperature distribution of z = 11 mm through the blood vessel. It can be seen from Figure 3(c) that the temperature of blood vessel is significantly higher than the temperature of gray matter, because the absorption coefficient of blood vessel is much greater than that of gray matter. Therefore, the blood vessel absorbs a large number of photons which causes the blood vessel to heat up significantly. The light energy is almost evenly distributed in the gray matter because the absorption coefficient of the gray matter is relatively small and the penetration depth of the laser is deeper. Therefore, the temperature distribution of the gray matter is relatively uniform. And because the light energy absorbed near the laser source is greater than that absorbed away from the laser source by the blood vessel, the temperature of the blood vessel near the laser source is higher than that away from the laser source. As can be seen from Figure 3(d), the temperature in the center of the blood vessel is lower than that on both ends of the blood vessel, because the light energy absorbed by the central part of the blood vessel is less than that absorbed by the both ends of the blood vessel.

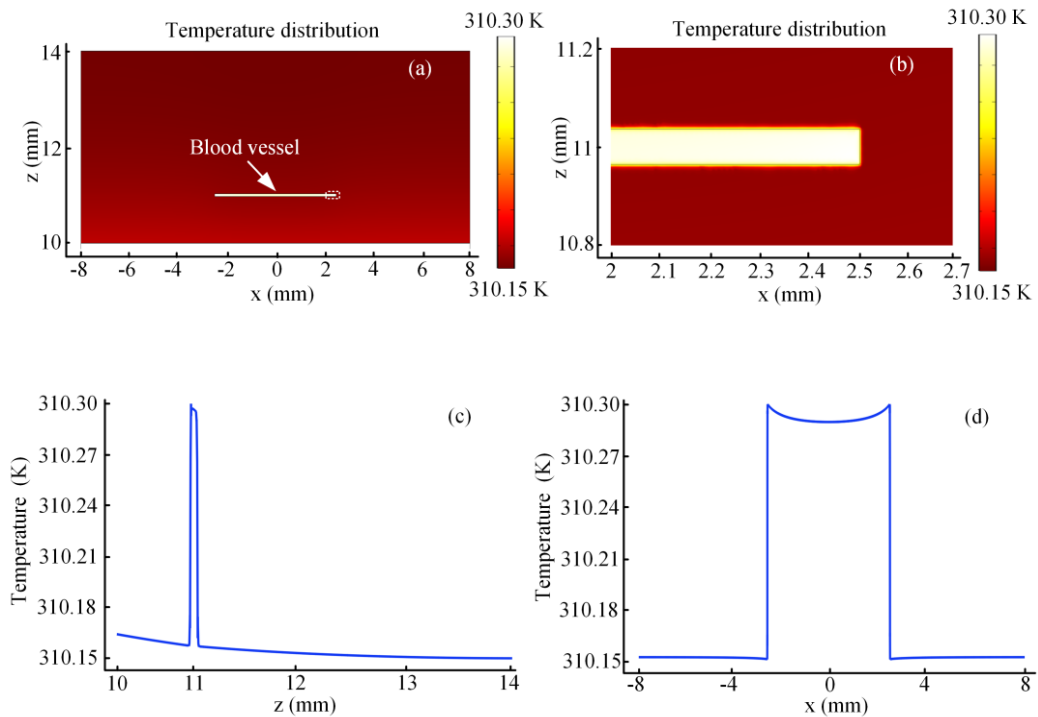

Figure 3. (a) shows the temperature distribution of the gray matter and blood vessel. The partial enlarged view of the temperature distribution of the gray matter and blood vessel is shown in (b). The temperature distribution of x = 0 mm

through the blood vessel and gray matter is shown in (c). (d) shows the temperature distribution of z = 11 mm through the gray matter and blood vessel.

## 4. CONCLUSION

We proposed a simulation model to study the interaction between laser and brain tissue based on COMSOL. The laser point source is located in the middle of the layer of water above the brain tissue and irradiates the brain tissue. The propagation of light in the brain tissue was simulated by solving the diffusion equation. And the temperature changes of gray matter and blood vessel were achieved by solving the biological heat transfer equation. The simulation results show that the remaining light energy of the blood vessel in the cerebral cortex is ~ 85.8% of the remaining light energy in the surrounding gray matter, and the temperature of blood vessel is 0.15 K higher than that of gray matter, and the temperature of gray matter hardly changes. This research has certain theoretical guiding for the optical imaging of the brain.